\documentclass{IEEEtran}
\usepackage{cite}
\usepackage{amsmath,amssymb,amsfonts}
\usepackage{graphicx}
\usepackage{textcomp,nicefrac}
\usepackage{siunitx}

\DeclareSIUnit{\sample}{S}
\DeclareSIUnit{\MSps}{\mega\sample\per\second}
\DeclareSIUnit{\GSps}{\giga\sample\per\second}

\DeclareSIUnit{\kHz}{\kilo\hertz}
\DeclareSIUnit{\MHz}{\mega\hertz}

\DeclareSIUnit{\bit}{b}
\DeclareSIUnit{\byte}{B}
\DeclareSIUnit{\Mbps}{\mega\bit\per\second}
\DeclareSIUnit{\Gbps}{\giga\bit\per\second}
\DeclareSIUnit{\MBps}{\mega\byte\per\second}
\DeclareSIUnit{\mebi}{Mi}
\DeclareSIUnit{\MB}{\mega\byte}
\DeclareSIUnit{\spill}{spill}

\def\BibTeX{{\rm B\kern-.05em{\sc i\kern-.025em b}\kern-.08em
T\kern-.1667em\lower.7ex\hbox{E}\kern-.125emX}}
\begin{document}
\title{Data acquisition system in Run-0a for the J-PARC E16 experiment}
\author{
    Tomonori Takahashi, 
    Kazuya Aoki, 
    Sakiko Ashikaga, 
    Wen-Chen Chang, 
    Eitaro Hamada, 
    Ryotaro Honda, 
    \\
    Masaya Ichikawa, 
    Masahiro Ikeno, 
    Shunsuke Kajikawa, 
    Koki Kanno,
    Daisuke Kawama, 
    Takehito Kondo,  
    \\
    Che-Sheng Lin, 
    Chih-Hsun Lin, 
    Yuhei Morino, 
    Tomoki Murakami,  
    Wataru Nakai, 
    Satomi Nakasuga, 
    \\
    Megumi Naruki, 
    Yuki Obara, 
    Kyoichiro Ozawa, 
    Hiroyuki Sako, 
    Susumu Sato, 
    Hiroshi Sendai, 
    Kazuki Suzuki, 
    \\
    Yudai Takaura,  
    Manobu Tanaka, 
    Tomohisa Uchida, 
    and 
    Satoshi Yokkaichi
\thanks{
    We would like to express our gratitude to 
    the staff members of J-PARC Hadron Experimental Facility 
    for their effort to construct and operate the J-PARC high-momentum beam line. 
    We also thank to 
    KEK electronics system group 
    and 
    open source consortium of instrumentation (OpenIt) 
    for their help in the development and test of the ASICs and PCBs. 
    We acknowledge the supports of 
    the Belle II collcaboration for UT3 and FTSW 
    and 
    CERN RD51 collaboration for SRS. 
    This work was supported by 
        the RIKEN SPDR program,
        Grant-in-Aid for JSPS Fellows 
            12J01196,  
            14J08572   
            and 
            18J20494,  
        and 
        MEXT/JSPS KAKENHI Grant numbers
            19654036,  
            19340075,  
            21105004,  
            26247048,  
            15H05449,  
            15K17669   
            and
            18H05235,  
        and
        the Ministry of Science and Technology of Taiwan Grant number 
        MOST108-2112-M-001-020. 
}
\thanks{
    T. N. Takahashi, 
    K. Kanno,
    W. Nakai, 
    and 
    S. Yokkaichi are with 
    Nishina Center for Accelerator-based Science, RIKEN (RNC), Wako, Saitama 351-0198, Japan 
    (e-mail: 
    tomonori@riken.jp; 
    kouki.kanno@riken.jp; 
    wataru.nakai@riken.jp; 
    yokkaich@riken.jp).}
\thanks{
    D. Kawama 
    was with RNC, Wako, Saitama 351-0198, Japan}
\thanks{
    K. Aoki, 
    E. Hamada, 
    M. Ikeno,
    K. Ozawa, 
    Y. Morino, 
    H. Sendai,            
    and 
    M. Tanaka
    are with 
    Institute of Particle and Nuclear Studies, 
    High Energy Accelerator Research Organization (IPNS KEK), Tsukuba, Ibaraki 305-0801, Japan 
    (e-mail: 
    kazuya.aoki@kek.jp; 
    ehamada@post.kek.jp; 
    ikeno@post.kek.jp; 
    ozawa@post.kek.jp;
    ymorino@post.kek.jp; 
    hiroshi.sendai@kek.jp; 
    tanakam@post.kek.jp).}
\thanks{
    T. Uchida
    was with IPNS KEK, Tsukuba, Ibaraki 305-0801, Japan}
\thanks{
    S. Ashikaga, 
    M. Ichikawa, 
    S. Nakasuga, 
    M. Naruki, 
    K. N. Suzuki, 
    and 
    Y. Takaura 
    are with 
    Department of Physics, Kyoto University, Sakyo-ku, Kyoto 606-8502, Japan 
    (email: 
    ashikaga@nh.scphys.kyoto-u.ac.jp; 
    m.ichikawa@nh.scphys.kyoto-u.ac.jp; 
    nakasuga.satomi.33a@st.kyoto-u.ac.jp; 
    m.naruki@scphys.kyoto-u.ac.jp; 
    suzuki@nh.scphys.kyoto-u.ac.jp; 
    takaura.yudai.37m@st.kyoto-u.ac.jp).}
\thanks{
    M. Ichikawa 
    is with Riken Cluster for Pioneering Research, RIKEN, Wako, Saitama 351-0198, Japan}
\thanks{
    W. C. Chang, 
    C. S. Lin, 
    and 
    C. H. Lin 
    are with 
    Institute of Physics, Academia Sinica, Nankang, Taipei 11529, Taiwan 
    (e-mail: 
    changwc@phys.sinica.edu.tw; 
    zslin@gate.sinica.edu.tw; 
    chihhsun.lin@phys.sinica.edu.tw).}
\thanks{
    R. Honda
    was with Department of Physics, Tohoku University, Sendai, Miyagi 980-8578, Japan. 
    He is now with IPNS KEK Tsukuba, Ibaraki 305-0801, Japan 
    (e-mail: 
    rhonda@post.kek.jp).}
\thanks{
    S. Kajikawa 
    is with Department of Physics, Tohoku University, Sendai, Miyagi, 980-8578, Japan.
    (e-mail:
    kajikawa@lambda.phys.tohoku.ac.jp).}
\thanks{
    T. Kondo 
    is with Department of Physical Science, Hiroshima University, Hiroshima, Hiroshima 739-8526, Japan. 
    (e-mail: 
    kondo@quark.hiroshima-u.ac.jp)}
\thanks{
    T. Murakami 
    is with Department of Physics, The University of Tokyo, Bunkyo-ku, Tokyo 113-8654, Japan.
    (e-mail: 
    mtomoki@post.kek.jp)}
\thanks{
    Y. Obara
    was with Department of Physics, The University of Tokyo, Bunkyo-ku, Tokyo 113-8654, Japan.}
\thanks{
    H. Sako, 
    and 
    S. Sato 
    are with Japan Atomic Energy Agency, Tokai-mura, Ibaraki 319-1195, Japan.
    (e-mail:
    sako@post.j-parc.jp; 
    susumu.sato@j-parc.jp)}
}

\maketitle

\begin{abstract}
    J-PARC E16 is an experiment to examine the origin of hadron mass through a systematic measurement of spectral changes of vector mesons in nuclei. 
    The measurement of \boldmath{$e^{+}e^{-}$} pairs from the decay of vector mesons will provide the information of the partial restoration of the chiral symmetry in a normal nuclear density. 
    To resolve a pulse pile-up and achieve good discrimination of \boldmath{$e^{\pm}$} from the background of a reaction rate of an order of 10 MHz,  
    the data acquisition (DAQ) system uses waveform sampling chips of APV25 and DRS4. 
    The trigger rate and data rate are expected to be 1 kHz and 130--330 MiB/s, respectively.  

    The DAQ system for readout of APV25 and DRS4 were developed, where events were synchronized by common trigger and tag data.        
    The first commissioning in beam, called Run-0a, was performed in June 2020 with about 1/4 of the designed setup. 
    The DAQ worked with a trigger rate of 300 Hz in the Run-0a and the main bottleneck was a large data size of APV25. 
    Further optimization of the DAQ system will improve the performance. 
\end{abstract}


\section{Introduction}
\label{sec:introduction}
\IEEEPARstart{J}{-PARC} E16 experiment\cite{pub:Yokkaichi2006} aims to investigate the origin of hadron mass through a systematic measurement of spectral changes of vector mesons in nuclei. 
The measurement will be performed at the high-momentum beam line, which is a newly constructed beam line, at J-PARC Hadron Experimental Facility with a new spectrometer.
A 30 GeV proton beam with an intensity of $1\times10^{10}$ protons per spill (2-seconds spill per 5.2 seconds cycle) irradiates thin targets. 
Proton-nucleus ($pA$) reaction occurs at an order of \SI{10}{\MHz} and produces vector mesons ($\rho/\omega/\phi$). 
If the vector mesons decay outside the nuclei, the measured mass shows the same value in the vacuum. 
On the other hand, if the vector mesons decay inside the nuclei (i.e. in medium), 
the mass spectra are expected to be modified due to an effect of the partial restoration of chiral symmetry with a finite baryon density.  

\begin{figure}[htbp]
    \centerline{\includegraphics[width=1.0\linewidth]{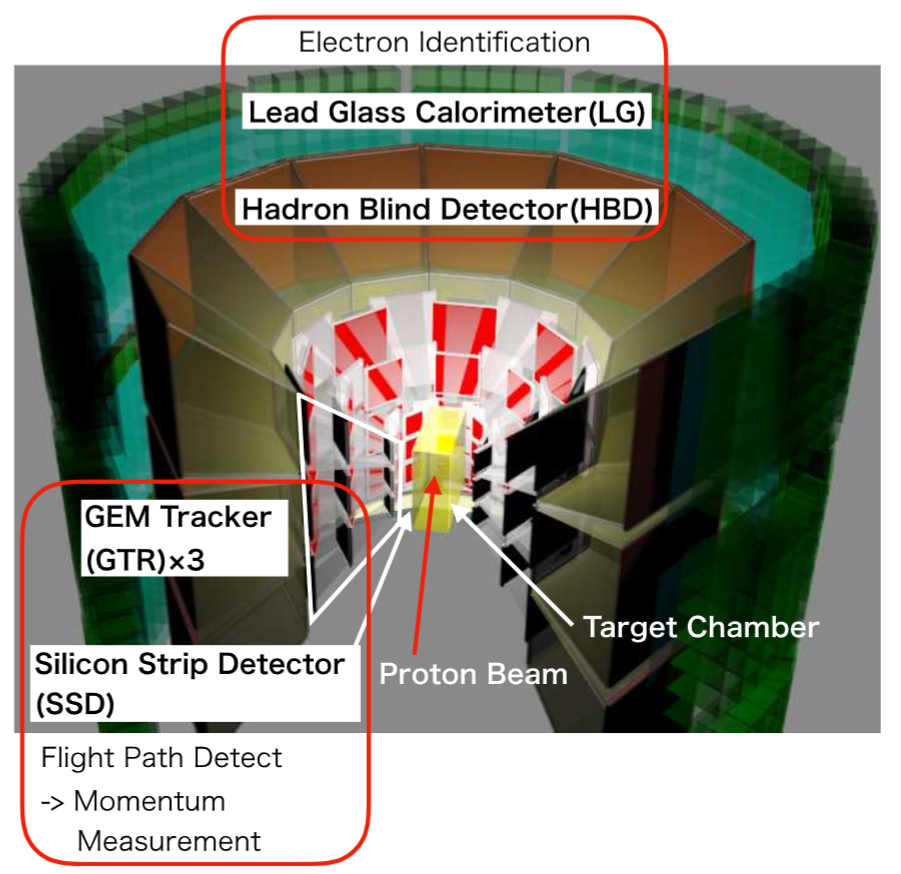}}
    \caption{Schematic view of the E16 spectrometer system (26 modules).}
    \label{fig:spectrometer}
\end{figure}

E16 focuses on the detection of the $e^{+}e^{-}$ pairs from the decay of $\rho/\omega/\phi$ mesons to avoid the final state interactions.
Therefore, the detector system was designed to measure $e^{\pm}$ precisely in a high-rate environment. 
The spectrometer is composed of 
        silicon strip detectors (SSD) 
        and  
        GEM trackers (GTR) 
    for momentum reconstruction of charged particles,
    and 
        hadron blind detectors (HBD)  
        and 
        leadglass calorimeters (LG)
    for electron identification. 
\figurename~\ref{fig:spectrometer} shows the schematic view of the spectrometer,  
which is placed inside a dipole magnet. 
The detector components are modularized so that  
each module covers \SI{30}{\degree} of horizontal and vertical angles. 
One module contains 
SSD (768 channels), 
three different sizes of GTR
( 
GTR1: \SI[product-units=power]{100 x 100}{\mm}, 432 channels, 
GTR2: \SI[product-units=power]{200 x 200}{\mm}, 720 channels, 
GTR3: \SI[product-units=power]{200 x 200}{\mm}, 1,080 channels, 
), 
HBD (1,538 channels), 
and LG (38 or 42 channels).
Table~\ref{table:detector} shows the individual number of readout channels for each detector 
and there are more than 110,000 in total. 

\begin{table}
    \caption{E16 detectors, readout devices and number channels}
    \label{table:detector}
\begin{tabular}{|l|r|c|c|r|}
    \hline
    Detector &  
    \begin{tabular}{r} Num. of\\ channels\end{tabular} & 
    readout chip & 
    digitizer & 
    \begin{tabular}{r} Num. of\\ trigger ch.\end{tabular}\\
    \hline
    SSD      & 19,968             & APV25   & APVDAQ   & \\
    GTR1     & 11,232             & APV25   & SRS-ATCA & \\
    GTR2     & 18,720             & APV25   & SRS-ATCA & \\
    GTR3     & 28,080             & APV25   & SRS-ATCA & 624 \\
    HBD      & 39,936             & APV25   & SRS-ATCA & 936 \\
    LG       & 1,060              & DRS4    & DRS4QDC  & 1,060 \\
    \hline
\end{tabular}
\end{table}

\section{DAQ design}
\label{sec:DAQ-design}
\begin{figure}[htbp]
    \centerline{\includegraphics[width=1.0\linewidth]{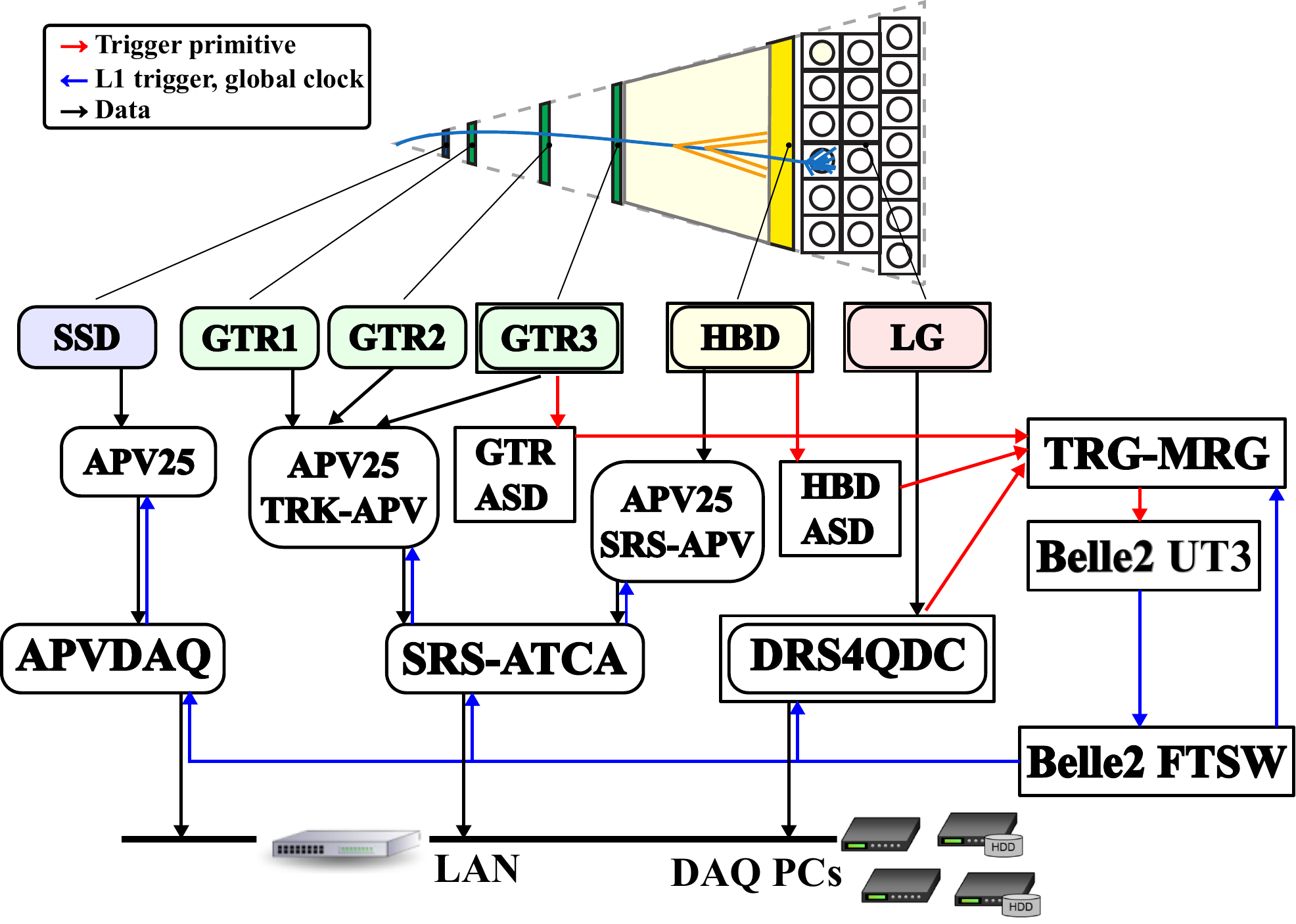}}
    \caption{E16 DAQ overview.}
    \label{fig:daq-overview}
\end{figure}

An overview of our data acquisition (DAQ) system is shown in \figurename~\ref{fig:daq-overview}.  
To resolve the pulse pile-up and 
to achieve the good discrimination of the background in the offline analysis, 
waveforms of the readout channels are recorded by using analog memory ASICs of APV25\cite{pub:Raymond2000} for SSD, 
GTR and HBD, and DRS4\cite{pub:Ritt2010} for LG, respectively.  
The sampling speed of APV25 and DRS4 are $\sim$\SI[per-mode=symbol,per-symbol=/]{40}{\MSps} and $\sim$\SI[per-mode=symbol,per-symbol=/]{1}{\GSps}, respectively.
In E16 DAQ, there are two types of front-end module for APV25. 
The first one is APVDAQ\cite{pub:APVDAQ}, which is used for SSD. 
The second one is SRS-ATCA\cite{pub:eicSys} for GTR and HBD. 
The front-end module for LG is called DRS4QDC. 
SRS-ATCA and DRS4QDC will be described in the following sections.  

These ASICs have neither a trigger output capability nor a continuous readout feature. 
To generate a trigger signal, LG and the last amplification layer of GEM foil of GTR3 and HBD are used as the source of the trigger primitive signals\cite{pub:Obara2015}. 
The read out is triggered at \SI{1}{\kHz} and the data rate of $\sim$\SI[per-mode=symbol,per-symbol=/]{660}{\mebi\byte\per\spill}
 (\SI[per-mode=symbol,per-symbol=/]{330}{\mebi\byte\per\second} in the \SI{2}{\second} beam duration, 
 and an averaged value is \SI[per-mode=symbol,per-symbol=/]{130}{\mebi\byte\per\second} over \SI{5.2}{\second} beam cycle) 
 is expected when zero-suppression in FPGA is applied. 

\section{Trigger system and event synchronization}
\label{sec:trigger-system}
The trigger primitive signals from GTR3, HBD and LG are digitized and merged by Trigger Merger boards (TRG-MRG\cite{pub:Ichikawa2019}), 
and afterwards sent to the trigger decision module (Belle2-UT3\cite{pub:Iwasaki2011}) via high speed optical links.  
The number of trigger primitive channels are listed in Table~\ref{table:detector}.  
UT3 generates several types of trigger data:  
    L1 trigger (physics trigger),  
    notification of spill-start or spill-stop, 
    calibration trigger, 
    sampling of hit-number counts and single event upset (SEU) counts, 
    which are encoded in 4 bits. 
The trigger data and associated tag data are distributed over standard CAT-7 LAN cables and repeater modules, Belle-2 FTSW\cite{pub:Nakao2012}.  
The busy signals from the front-end modules are transferred by the same LAN cable and collected.  
Although the serial data are encoded in B2TT protocol, 
the distributed clock frequency and line rate are not same as the Belle-2 DAQ, which uses a \SI{127}{\MHz} clock.  
It is derived from the RF frequency of the SuperKEKB and there is no advantage in using it in the other site.   
Instead, a \SI{125}{\MHz} clock and \SI[per-mode=symbol,per-symbol=p]{250}{\Mbps} line rate are adopted. 
The unique tag data consists of 
    48-bit timestamp (LSB=\SI{8}{\nano\second}), 
    16-bit spill ID 
    and 32-bit event ID.  
The received tag data are checked to detect any event synchronization error.       

Only SRS-ATCA can be directly connected to FTSW. 
RPV-260\cite{pub:RPV-260} was designed to play as an interface bridge between RJ45 and KEK-VME backplane bus\cite{pub:Igarashi2010}. 
For APVDAQ, a trigger interface module (TIM) was developed. 
TIM is also used as the clock fan-out module to synchronize APVDAQ. 

\begin{figure}[htbp]
    \centerline{\includegraphics[width=1.0\linewidth]{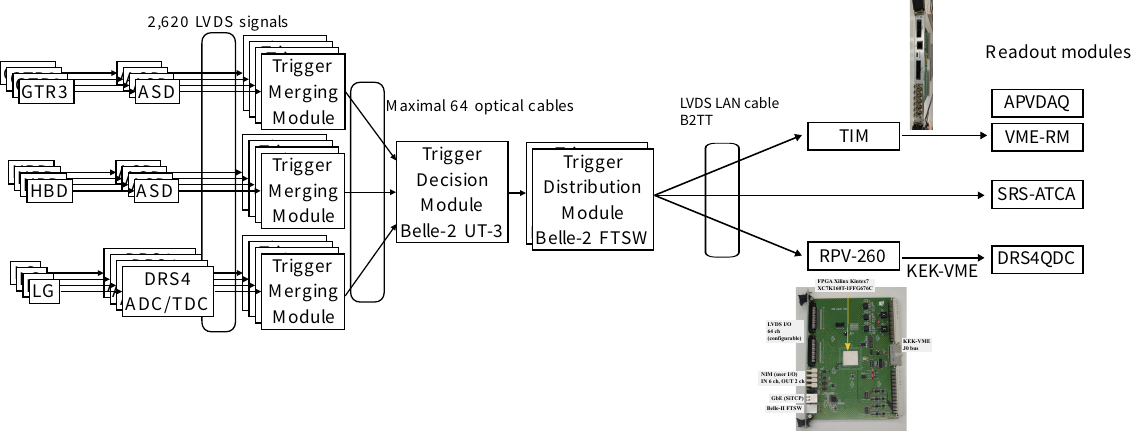}}
    \caption{Schematic view of trigger data flow.}
    \label{fig:trigger-data-flow}
\end{figure}

\section{Front-end module for GTR and HBD}
\label{sec:FEM-GTR-HBD}
\begin{figure}[htbp]
    \centerline{\includegraphics[width=1.0\linewidth]{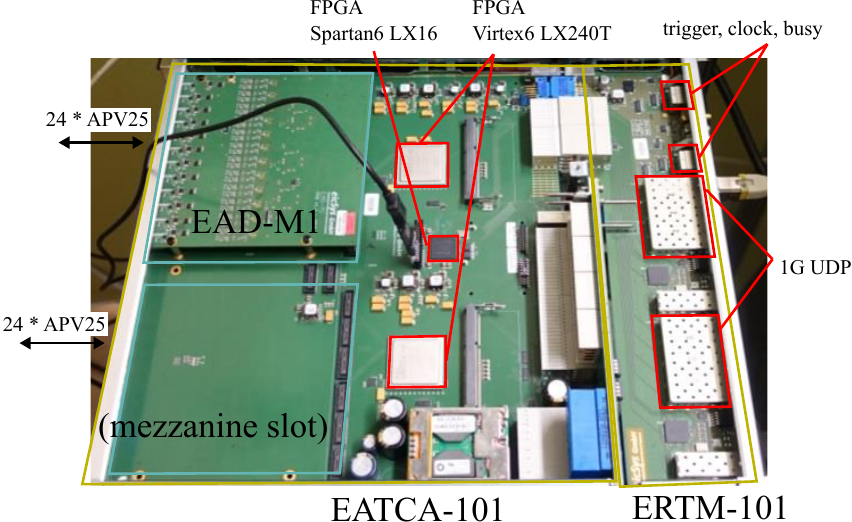}}
    \caption{Photo of SRS-ATCA module.}
    \label{fig:srs-atca}
\end{figure}

SRS-ATCA, 
a variant of CERN-RD51\cite{pub:Roplewski2008}'s Scalable Readout System\cite{pub:Martoiu2013}, 
was adopted as the front-end module for GTR and HBD. 
A photo of SRS-ATCA is shown in \figurename~\ref{fig:srs-atca}.  
The SRS-ATCA hardware was designed by a German company eicSys. 
The module consists of 
    an ATCA blade (EATCA-101), 
    two mezzanine cards (EAD-M1) 
    and one RTM (ERTM-101). 
Two main FPGAs of Xilinx Virtex-6 LXT240 on the blade read the ADC data and manage each mezzanine card and APV25 chips, 
and one sub FPGA of Xilinx Spartan-6  
LX16 configures the main FPGAs. 
Front-end hybrid cards of APV25 and the mezzanine cards are connected by HDMI cables. 
Each mezzanine card has three 12-bit 8-channel ADCs (Texas Instruments ADS5281)
 to cope with 24 APV25 chips. 
Therefore, one SRS-ATCA module can handle 
\SI{128 x 24 x 2}{} = 6,144 channels in total.  
The trigger and tag data, which are encoded in B2TT protocol, are received via the RTM. 
The data of main FPGAs are transferred to PC by UDP.

The firmware of SRS-ATCA has been extended for E16 as follows.
Front-end modules are located near the spectrometer, 
where radiation level is not negligible.  
Therefore, soft error mitigation (SEM) core was implemented by using Xilinx IP to reduce the SEU failure. 
In addition, a watchdog timer is implemented to watch the loss of heartbeat from the SEM core. 
When the heartbeat from the SEM core is lost or the detected error is uncorrectable, 
the SEM core takes a control of internal configuration access port (ICAP) and reload the firmware from the on-board SPI-ROM to reboot the FPGA.  
The number of samples per event was chosen to be 24 because the GTR has a long drift time to obtain good position resolution. 
The APV25 and ADC are synchronized at \SI{41.67}{\MHz} which is derived clock from the global clock of \SI{125}{\MHz}.  
TDC (LSB=\SI{2}{\nano\second}) is implemented to measure the time difference between the global clock of \SI{125}{\MHz} and the sampling clock of \SI{41.67}{\MHz}. 
The TDC uses a simple counter running at \SI{500}{\MHz} clock. 

\section{Front-end module for LG}
\label{sec:FEM-LG}
A waveform digitizer with DRS4, DRS4QDC, was developed for the LG readout. 
The board has 16 analog input channels. 
The input signals are split and fed into DRS4 chips and comparators.  
The former capture the waveform at \SI[per-mode=symbol,per-symbol=/]{960}{\MSps} and the latter generate trigger primitives.  
The board has four DRS4 chips, where  
a pair of capacitor arrays of DRS4 are cascaded to extend the Level-1 trigger latency up to \SI{2}{\micro\second}. 
To reduce the data size, only 200 samples in the region of interest (RoI) for each trigger are captured.
DRS4 output is digitized by 12-bit 8-channel ADC, Analog Devices AD9367, which is used with a \SI{30}{\MHz} sampling frequency.  
These chips are managed by a Xilinx Spartan-6 LX150 FPGA.  
A FPGA-based TCP/IP processor, SiTCP\cite{pub:Uchida2008}, where 100BASE-T is used for the data transfer.
Although 100BASE-T is enough to send event frames at the data rate of \SI[per-mode=symbol,per-symbol=/]{6.4}{\MB\per\second}, corresponding to the \SI{1}{\kHz} L1 trigger without zero-suppression (200 sample points $\times$2 Bytes $\times$ 16 channels $\times$ \SI{1}{\kHz}), 
the functionalities of the waveform correction and data size reduction are implemented in the FPGA. 
First, 
the offset variation of raw waveform data is corrected with a block-RAM based lookup table. 
Next, the sequential order of the cascaded-pair data is sorted into a correct order. 
Then, so-called a symmetric spike noise, which appears around the last capacitor cell in the RoI of the previous trigger, is removed by linear interpolation using the values of neighbor cells of the spike. 
Next, the sum of the waveform within a specified length is calculated and compared with a threshold to decide whether the channel has a hit or not.  
Finally, the waveform data of the hit channel is further reduced by a delta-compression. 
It is a loss-less compression and the difference of the amplitude between the next sample point is calculated and classified to determine the bit width of the packed data structure. 

DRS4QDC receives the trigger and tag data via the KEK-VME J0 backplane bus. 
The KEK-VME J0 bus has 7 differential lines to share the trigger and tag data\cite{pub:Igarashi2010}. 
The original J0-protocol uses the bus as an asynchronous parallel data bus and doesn't send more than 5-bit data. 
However, DRS4QDC uses it as a source-synchronous bus 
to convey 4-bit trigger data and 96-bit tag data, 
whereas a backward compatibility is kept. 
The SEM is also implemented in the same way as SRS-ATCA except for the difference of the FPGA family. 

\begin{figure}[htbp]
    \centerline{\includegraphics[width=1.0\linewidth]{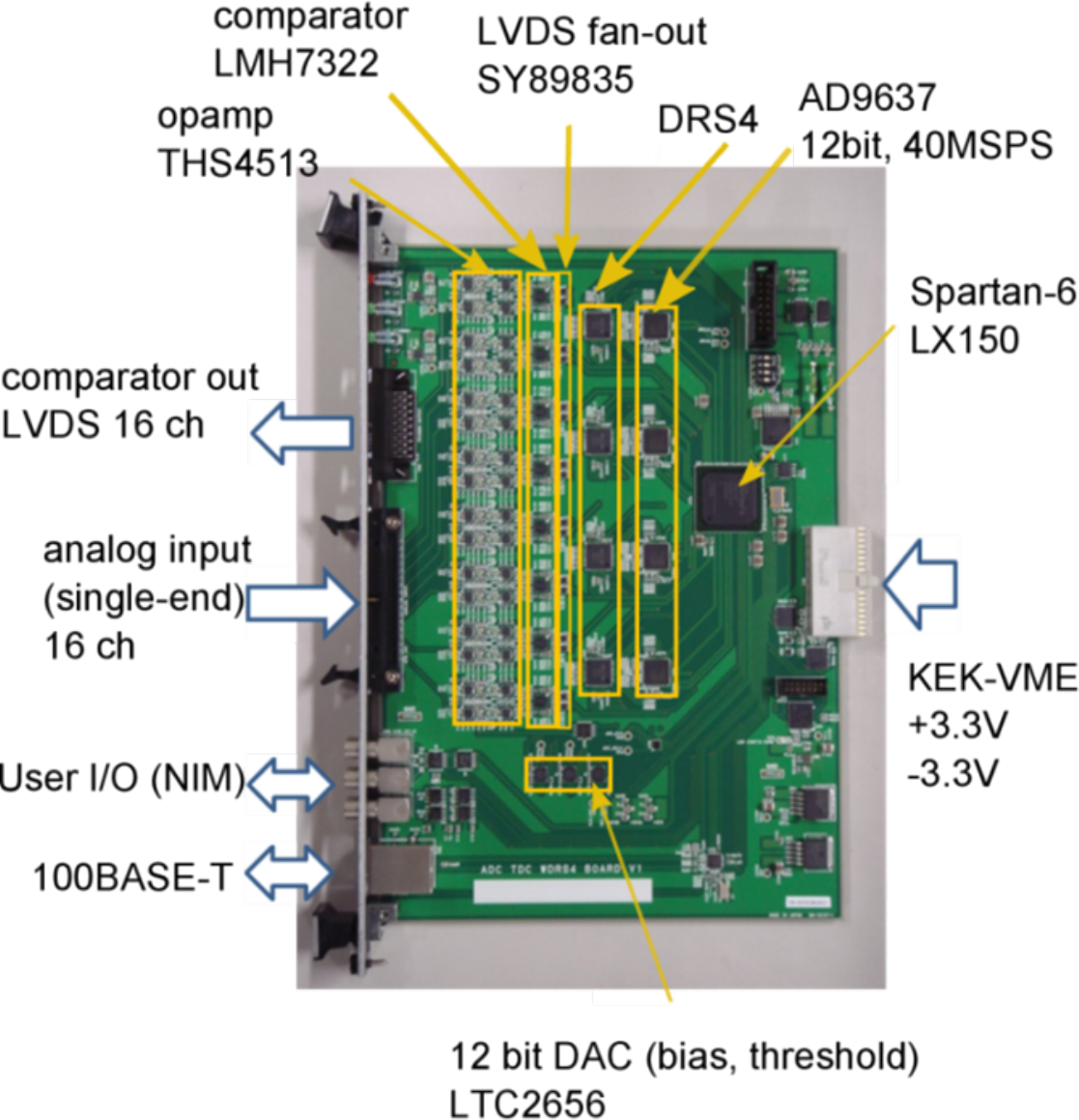}}
    \caption{Photo of DRS4QDC module.}
    \label{fig:drs4qdc}
\end{figure}

\section{Beam commissioning}
\label{sec:commissioning}
\begin{figure}[htbp]
    \centerline{\includegraphics[width=1.0\linewidth]{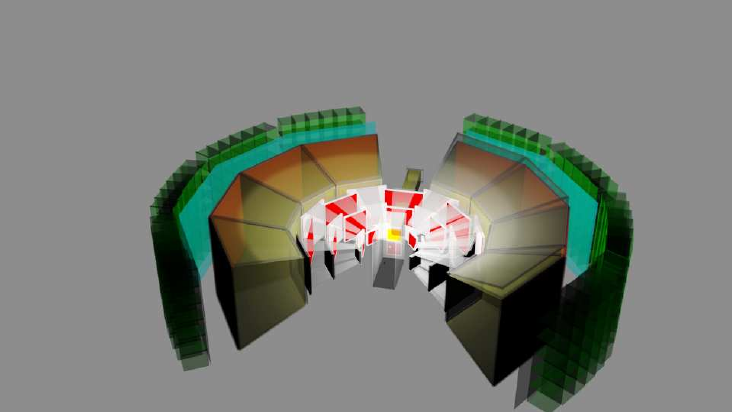}}
    \caption{E16 spectrometer setup in Run-1 (8 modules in the middle part of the full setup). 
             The setup of Run-0a (Run-0b) is composed of 6(6) SSD, 6(8) GTR, 4(6) HBD, and 6(6) LG.}
    \label{fig:run1}
\end{figure}

E16 takes a staging approach in the spectrometer construction. 
Following stages are planned.  
The first commissioning of detectors and DAQ with beam, called Run-0a, was carried out in June 2020, 
where about 1/4 of full detectors (namely, 6 SSD, 6 GTR, 4 HBD, and 6 LG) were installed. 
The second commissioning (Run-0b), is scheduled in January 2021, 
where 6 SSD, 8 GTR, 6 HBD, and 6 LG will be used. 
The first physics run with 8 modules is planned in 2022 (Run-1), 
whereas the experiment with the full-equipped spectrometer will be performed within a few years (Run-2).   
Schematic views of Run-1 and Run-2 are shown in 
    \figurename~\ref{fig:run1}
    and \figurename~\ref{fig:spectrometer}, 
    respectively. 

In Run-0a, an online event building software was not ready. 
Thus, data of the front-end modules were collected by separate processes for each module and written into the hard disks,  
while the trigger and tag were common to them. 
One PC
with two Intel Xeon E5-2630 v4 10-core processors, 
DDR4 256 GB RAM, 
a network card of Intel X710D4 (\SI[per-mode=symbol,per-symbol=p]{10}{\Gbps}$\times 4$, only 22ports were enabled), 
and CentOS 7 operating system 
was used for the data collection of 4 SRS-ATCA and 18 DRS4QDC. 
For readout of 12 APVDAQ, 
the J-PARC hadron DAQ software\cite{pub:Igarashi2010} was used, 
where two single-board computers (SBCs), XVB601 and XVB602, were installed and each SBC passed data of 6 APVDAQ to the recorder PC.  
A data merger software based on ZeroMQ (FairMQ\cite{pub:FairMQ}) was developed and used for TRG-MRG and UT3. 

The measured DAQ rate was approximately \SI{300}{\hertz}. 
The main bottleneck was that a long dead time of $\sim$\SI{3}{\milli\second} in SRS-ATCA,  
where no zero-suppression was applied in the FPGA.  

\section{Summary and outlook}
\label{sec:summary}
J-PARC E16 experiment has successfully commissioned in June 2020 to investigate the origin of the hadron mass.  
The DAQ system adopted waveform sampling with analog memory chips of APV25 and DRS4.  
The front-end module and/or the readout firmware of those chips were developed.   
During the first beam commissioning, about 1/4 of full-equipped setup was installed.  
The DAQ worked with a trigger rate of approximately \SI{300}{\hertz}, 
which was 1/3 of the design goal. 
It was caused by the long data transfer time of \SI{3}{\milli\second} in SRS-ATCA.  

Further developments are in progress. 
To improve the trigger accept rate, 
development of the zero suppression firmware as well as 
an FIR filter to compensate the signal distortion by a long transmission line are ongoing for SRS-ATCA. 
We also test parallel readout of APVDAQ by using MOCO\cite{pub:Baba2012} and an upgraded module. 
Moreover, the development of the DAQ online software for event-building and monitoring is to be done.  


\end{document}